\documentclass[reprint,
 aps,
]{revtex4-2}

\usepackage{url}

\usepackage[breaklinks]{hyperref}
\usepackage{breakurl}

\usepackage{float}
\usepackage{graphicx}
\usepackage{dcolumn}
\usepackage{bm}
\usepackage{lineno,hyperref}
\usepackage{graphicx}	
\usepackage{amsmath}	
\usepackage{amssymb}	
\usepackage{graphicx}
\usepackage{mathrsfs}
\usepackage{natbib}
\usepackage{color}
\usepackage{diagbox}
\usepackage[dvipsnames]{xcolor}
\usepackage{soul}

\begin{document}

\preprint{APS/123-QED}

\title{Does the Vacuum Gravitate on Microscopic Scales?  Rydberg Atoms Indicate Probably Not \\ \small (Accepted in PRD)}

\author{Suman Kumar Kundu}
 \email{skundu@syr.edu}

\author{Arnab Pradhan}%
\email{arpradha@syr.edu}

\author{Carl Rosenzweig}
\email{crosenzw@syr.edu}

 \affiliation{Department of Physics, Syracuse University,
Syracuse, NY 13244, USA
}%


\date{\today}

\begin{abstract}
\
The cosmological constant presents one of the most fascinating and confounding problems in physics.  A straightforward, seemingly robust prediction of quantum mechanics and general relativity is that the vacuum energy gravitates.  Therefore, the cosmological constant should be enormous. It is minuscule.  Since there is no understanding of why the cosmological constant is so small, it is important to test this idea in many different situations.  In particular, given the span of distances in astronomy and particle physics, it is vital to test the gravitation of vacuum energy on as many distance scales as we can.  Rydberg atoms open up a new set of distances for exploration. It is satisfying to measure the cosmological constant with an atom, but its main significance is extending measurements to microscopic distances. Here, too, there is no evidence of the gravitation of the vacuum.  At scales of a micron and less, we place a limit of $7$ GeV on the scale of gravitating vacuum energy, well below the scale of $100$ GeV of the SM of particle physics.
\end{abstract}

\maketitle

 \section{Introduction}
One of the most pressing and perplexing problems facing theoretical physics is the explanation of the observed cosmological constant (CC) \cite{Einstein17,Zel67,Weinberg00,Martin:2012bt}. A suggestive, back-of-the-envelope calculation leads to an estimate off by 120 orders of magnitude! The calculation is motivated by the most basic features of quantum mechanics and general relativity. Quantum mechanics tells us that there is a zero-point energy, while quantum field theory tells us that the vacuum energy (VE) is pervasive and huge. Energy is important in physics, but only energy differences enter most physical situations. This changes in general relativity, where the absolute value of energy participates in gravity.  Since the zero-point energy $\hbar \omega/2$ is ubiquitous in quantum field theory, we have to sum this contribution over all modes of the field. The contribution to the VE density from a field of mass $m$, incorporating a momentum cut-off $\Lambda$ is (unless otherwise stated, we use units with $c=\hbar =1$)

\begin{equation}
     \rho_{v} = \int^{\Lambda} \frac{d^3p}{(2 \pi)^3} \frac{1}{2} \sqrt{(p^2+m^2)}=\frac{\Lambda^4}{16 \pi^2}
     \label{eq:qve}
\end{equation}
Thus, the vacuum acquires energy, and this energy should be a source of gravity.  The gravitational interaction of the VE in turn gives rise to a CC. Since we do not have a quantum theory of gravity, the Planck scale ($M_P$) represents the limit of our knowledge and is a natural candidate for the cutoff in eqn. \ref{eq:qve}, giving  $\rho_{v}$ of order $(M_P)^4$, off by record $120$ orders of magnitude from the measured value of $\rho_{v}$, $(0.002  \ \text{eV})^4$ \cite{Planck18}.

If we are careful about introducing the cutoff,  $\rho_{v}$ is determined by the largest mass scale in the theory \cite{Akhmedov02,Koksma11}. 
 We denote this scale by $\Lambda_{\rm{UV}}$ and define the associated VE as
\begin{equation}
\rho_v=\Lambda_{\rm{UV}}^4
\label{eq:rhovdef}
\end{equation}
The Standard Model (SM), outstandingly successful in describing physics from radioactive decay to the Higgs particle, has a typical scale of $100 \ \text{GeV}$.

Barring incredible fine-tuning and/or theoretical gymnastics, the tiny value of the CC is evidence that the vacuum does not gravitate on cosmological scales. Explaining this is a challenge for theoretical physics. Crucial to this picture is the conviction that the VE gravitates. But is this true? Since this problem was first noticed by Pauli almost a century ago \cite{Straumann02}, hundreds, if not thousands, of articles have appeared searching for a solution (for a review, see \cite{Copeland06,Burgess13,Polchinski06}). Some were clever, some ingenious, some promising, but none compelling. Some completely decouple VE from gravity, while others partially decouple.  Numerous models accept the reality of gravitating VE, while providing a mechanism to vitiate its effects on the scale of our current cosmos. 
Some models involve the incoherent addition \cite{Zwane18,Hogan04,Wang17,Wang20,Bengochea20,Firouzjahi22,KPR22}, and attendant dilution, over long distances or time scales of patches of basic VE. Others rely on damping from long times and large distances (see \citet{Brax17} for a review and further references). 

The cosmological constant remains a mystery.  Some proposed solutions require violation of cherished principles such as Lorentz invariance and / or locality \cite{Csaki01,Arkani02}. This desperation is a signal that we are missing something crucial.  To some extent, the problem rests on the measurement of a single number. Therefore, it is vital to explore other possible measurements.  VE as a source of the CC rests on theories that span distances from nuclear size to our observable universe, a span of $42$ orders of magnitude. We cannot rule out the possibility that vacuum, classical, or quantum, manifests itself differently on micro-, macro-, or cosmic scales. The fact that many proposed solutions are consistent with, or even demand, that VE behaves differently at cosmic and non-cosmic distances motivates such a study \cite{Zwane18,Hogan04,Wang17,Wang20,Bengochea20,Firouzjahi22,KPR22}. There are even proposals that the CC is explicitly scale dependent  \cite{Bonanno21,Cuzinatto22}. In Section \ref{eftsec} we examine how effective field theories suggest distance-dependent effects of VE.  

A further, heuristic, argument for experiments to probe dark energy (for which a CC is a prime candidate) at non-cosmological scales is that dark energy may behave like dark matter--uniform on cosmic scales but clumpy at smaller scales. This was a prime motivation for the proposals of Perl \cite{Perl10,Adler11} to justify a new class of experiments measuring the gravitating effect of dark energy at terrestrial distances.  It is worthwhile to concentrate on simple, direct implications of gravitating VE and check what evidence there is for or against the basic idea behind the problem of the CC. This, after all, is the major job of experimental physics, to explore uncharted waters.  Cosmic distances offer the most stringent constraints on the energy scale of gravitating VE, which is found to have a value $10^{-3}$ eV. As we probe ever shorter distances from galactic to astrophysical to laboratory (terrestrial) data, the experimental upper limit on possible enrgy scales of gravitating VE grows, to $10$ eV on solar system scales (see \cite{Martin:2012bt}) and about $100$ keV on laboratory scales. This leaves distances below a millimeter to explore. Amazingly, precision atomic-physics experiments can explore exactly that regime, and hence can be added to the list of meaningful probes of the CC.  They can determine if the quantum vacuum gravitates at microscopic distances. In this paper, we examine the evidence for or against gravitating VE at these unexplored atomic scales. 

An atom immersed in a background VE experiences a perturbative gravitational interaction proportional to $r^2$ times $\rho_v$, the VE density. Highly excited Rydberg atoms (RA), with $r \propto n^2$, thus feel a perturbation proportional to $n^4$. The large values of $n$ available ($100$ or more) coupled with the incredible accuracy of the order of $10^{-10}$ eV attainable in modern Rydberg experiments \cite{Merkt16,Peper19} enable us to put a limit on the gravitating component of $\rho_{v}$ of $ (7 \ \text{GeV})^4$ at distance scales less than one micron. This sets an upper limit on the scale of a gravitating VE of $7 \ \text{GeV}$ on this length scale. Although not as good as astronomical or macro limits nor remotely competitive to the cosmological determination of the CC this is interesting because it eliminates most of the putative contributions, with characteristic energy of $100 \ \text{GeV}$, from particle physics, at hitherto unexplored microscopic distances. This has interesting implications for cosmology and nascent theories of quantum gravity. 
 
We find it remarkable that an atom in a terrestrial laboratory can teach us something about cosmology. If we had neither knowledge nor ability to make cosmological or astronomical measurements atomic physics, by itself, tells that we have a serious CC problem (A future, different terrestrial experiment has been proposed by \citet{Avino19}). More significant is that Rydberg experiments provide important information about a previously inaccessible region - microscopic distances of less than a micron - to test whether VE does or does not gravitate. Failure to see evidence of the expected gravitational effect of the vacuum at distances greatly different from cosmic or macro-scales indicates that the problem lies in the assumption that the vacuum gravitates. 

\section {What Atoms Know About Expansion of the Universe and VE} 
Consider a hydrogen atom in an expanding universe.   We adopt the Newtonian approach of \citet{Price12} since this keeps the physics simple, and can be justified with a full general relativistic treatment (see \citet{Carrera10} for a review and references). The physical position $r$ of the electron relative to its (heavy) nucleus is,
\begin{equation}
   r =a(t)R 
\end{equation}
where $R$ is the comoving coordinate and $a(t)$ the scale factor of the universe. If the universe expands exponentially because of a non-zero CC the scale factor is,
\begin{equation}
   a(t) \propto e^{\sqrt{\frac{\Lambda_{\text{CC}}}{3}}t}
\end{equation}where
\begin{equation}\label{ccdef}
\Lambda_{\text{CC}} =\frac{ 8\pi }{M_P^2} \rho_v 
\end {equation} 
The electron, due to expansion, is subjected to a repulsive force,
\begin{equation}
F_{exp}=m_e\frac{\Lambda_{\rm{CC}}r}{3}
\end{equation}
where $m_e$ is the mass of the electron. The equation of motion of the electron, including contributions from the Coulomb, centrifugal, and repulsive force, is
\begin{equation}
   m_e\frac{d^2r}{dt^2}-\frac{L^2}{m_e r^3}=-\frac{\alpha}{r^2}+\frac{m_e\Lambda_{\rm{CC}}r}{3}
    \label{eq:eom}
\end{equation}
where $\alpha$ is the fine structure constant and $L$ is the angular momentum (we use $m_e$ explicitly as opposed to \cite{Price12}). From eqn. \ref{eq:eom} we identify an effective potential,
\begin{equation}
    V_{eff}= \frac{L^2}{2m_e r^2}-\frac{\alpha}{r}- \frac{m_e \Lambda_\text{CC}}{6} r^2
\end{equation} 
\citet{Price12}  introduce two characteristic time scales, $T_{atom}$ the typical atomic orbital period and $T_{exp}$ the ``Hubble" expansion time. They study $V_{eff}$ for different ratios of $T_{atom} / T_{exp}$. For sufficiently small values of this ratio, the atom remains bound and resists the expansion of the universe, while for larger values ($>1/4$) it escapes and shares in the expansion. This is in accordance with the intuitive feeling that a tightly bound system decouples from expansion. The electron, however, ``remembers" the expansion since it now sits in a modified potential.  
Rather than the purely classical $V_{eff}$ we treat the electron quantum mechanically and use it as our modified potential in Schrodinger's equation
\begin{equation}
    V= -\frac{\alpha}{r} - m_e\frac{\Lambda_\text{CC}}{6}r^2 
    \label{eq:ep}
\end{equation}

RAs are capable of noticing this modification as a perturbation to the Coulomb potential, and thus can measure or place a limit on the CC!

In the above analysis, it is irrelevant whether the modification to the effective potential is due to a universal cosmological expansion or to the local effects of the VE density $\rho_{v}$. Except for approaches to the CC that exorcise VE entirely from gravitating \cite{Dadhich16, Boyle21,Brodsky22,Adler21,Lombriser19,Sundrum04,Dvali07,Arkani02} all other approaches see, e.g. \cite{Brax17,Padilla15} have some contribution to $V_{eff}$. The perturbation to the Coulomb potential is then  
\begin{equation}\label{pot}
    \delta V = - \frac{4\pi}{3} G m_e \rho_v r^2 = -\frac{4\pi}{3} \frac{ m_e \rho_v r^2}{M_p ^2 }
\end{equation}
We rewrite $G$ in terms of the Planck mass $M_P \sim 10^{19}$ GeV.  If, as we see in Section \ref{shift}, $\rho_v/M_P^2$ is small, $\delta V$ leads to a perturbation to the electron energy, $\delta E$.

As a check of the consistency of this viewpoint, namely that the electron decouples from expansion due to a local $\rho_{v}$  but retains a memory with a shifted potential, we compare $T_{atom} \sim 10^{-16}$ s for the ground state of a H atom to $T_{exp} \sim 10^{-6}$s, for $\rho_v= (100\ \text{GeV})^4$.  Furthermore, even after the electron decouples from the expansion and is excited to a high state (say $n = 100$), it is still well within the de-Sitter horizon $\sqrt{3/\Lambda_{\text{CC}}} \sim 10$ cm. The potential eqn. \ref{eq:ep} turns over only for Rydberg states with $n > 160$, so an excited Rydberg atom with $n \sim 100$ is safely bound.

\section{Brief Overview of Rydberg Experiments}
RAs are the template of some of the most precise experiments in physics,  We quote \citet{Merkt16} ``With the rapid development of methods to generate cold samples of molecules and the extension of frequency combs to shorter wavelengths, measurements of molecular....energies with sub-MHz precision are becoming possible by Rydberg-state spectroscopy.... Alkali metal atoms offer distinct advantages for precision measurements of....energies: their Rydberg states....can be reached by single-photon UV excitation from the ground state, i.e. in a range where modern frequency-metrology tools can be fully exploited.  Alkali metals can be easily laser-cooled to sub-mK temperatures, so that Doppler and transition-time broadening become almost negligible.  Finally, the closed-shell nature of the ion core implies that Rydberg series of alkali-metal atoms can be accurately treated as single ionization channels with the...Ritz formula."

Experiments on RAs are now capable of exciting the atoms to energy levels of the order $n = 100$ while measuring the energy levels with a precision of $10^{-10}$ eV. \citet{Peper19} performed such a precision experiment to measure the absolute frequency of transitions from the ground state to large $n$ Rydberg states. They first prepare sub-Doppler-cooled  $^{39}\text{K}$ samples in the $4s_{1/2}$  ground state by confining them inside magneto-optic traps. They excite the atoms with pulses of frequency tune-able light from the $4s_{1/2}$ state to over 20 different $np_{1/2}$ and $np_{3/2}$ Rydberg states.  In these experiments $n$ ranging from 22 to 100 are achieved. They record these Rydberg states by millimeter-wave spectroscopy with extraordinary precision.  We employ these high-precision energy measurements for $^{39}\text{K}$ atoms \cite{Peper19} to place a limit on the magnitude of the VE density. The data are accurately described, to a few parts to $10^{-6}$/cm, by the modified Ritz formula (in the following formulae, we use $h = 1$ to convert wave numbers to energies)
\begin{align}
\label{eq:ritz}
    E_{nlj} = E_I - \frac{R_{\text{K}}}{(n - \delta_{lj}(n))^2}
\end{align}
where
\begin{align}\label{defect}
    \delta_{lj}(n) &= \delta_{0,lj} + \frac{\delta_{2,lj}}{(n-\delta_{0,lj})^2} +  \frac{\delta_{4,lj}}{(n-\delta_{0,lj})^4}\nonumber\\ &\quad+  \frac{\delta_{6,lj}}{(n-\delta_{0,lj})^6} + ...
\end{align}
are the energy-dependent quantum defects for the respective series. $E_I$ is the ionization energy and $\text{R}_{\text{K}}$ is the reduced Rydberg constant for $^{39}\text{K}$.

Quantum defects $\delta$ were originally introduced by Ritz as purely phenomenological factors that provided an accurate representation of his data. Subsequently Sommerfeld, in old quantum theory \cite{Sommer20}, and then Hartree, using quantum mechanics \cite{Hartree28}, derived these quantum defects to account for the effects of the short range corrections (due to the inner electrons)  to the Coulomb potential experienced by the outer electron.  Hence, they are experimentally and theoretically well founded.

\section{Vacuum induced energy shifts}\label{shift}

 Consider an excited Rydberg atom immersed in a vacuum teeming with energy described by a constant VE density $\rho_v$. If the vacuum gravitates, the outer electron experiences a gravitational potential given by eqn. \ref{pot}. For a hydrogen-like atom, the shift in energy due to this potential is given by
 \begin{align}
    \delta E &= -\frac{4\pi m_e \rho_v \langle r^2\rangle}{3 M_P^2} \\
    &= -\frac{4 \pi m_e \rho_v}{3M_P^2} \frac{ n^2}{2\alpha^2 m_e^2}[5n^2 - 3l(l+1) + 1]\label{hydptb}
\end{align}
where $\langle r^2\rangle$ is the expectation value of $r^2$ in the hydrogen-like atom, $n$ is the principal quantum number, $l$ is the angular momentum quantum number, and $\alpha$ is the fine structure constant. We note that \cite{Gill87} has discussed the shift in energy levels of RA due to the tidal force experienced in strong gravitational fields. They were concerned with astrophysical effects while here, taking advantage of the increased precision of laboratory experiments we focus on atoms in the lab. The physical and mathematical basis of both studies is very similar. For $n>>l$, the leading-order energy perturbation in $n$ is
\begin{equation}
\label{eq:ptbformula}
    \delta E (n) \sim -\frac{10\pi}{3}\frac{\rho_{v}}{\alpha^2 M_P^2 m_e}n^4
\end{equation}

Since the perturbation $\delta E$ (eqn.\ref{eq:ptbformula}), shifts the energy levels, the shifts are more relevant to us than the absolute values of $E_I$. Therefore, we chose to fit the energy differences between the various levels $n > 22$ and $n = 22$. We use the $np_{3/2}$ data from \citet{Peper19} for the fit. Using the $np_{1/2}$ data leads to the same results. We chose $n = 22$ since the most precise data start from $n=22$ and all the $n$ are sufficiently high that the Ritz formula is accurate. Since the fits are relatively insensitive to defects beyond the second order in eqn. \ref{defect}, we fit the energy differences to the expression
\begin{equation}
\begin{split}
    E_{nlj} - E_{22lj} &= R_K\Bigg[\frac{1}{\left(22 - \delta_{0,lj}-\frac{\delta_{2,lj}}{(22-\delta_{0,lj})^2}\right)^2}\\&\quad-\frac{1}{\left(n - \delta_{0,lj}-\frac{\delta_{2,lj}}{(n-\delta_{0,lj})^2}\right)^2}\Bigg] \\&\qquad+ \left(\delta E (n) - \delta E (22)\right)
\end{split}
  \label{eq:ritzwp}
\end{equation}

\begin{figure}[h]
    \centering
    \includegraphics[width=0.48\textwidth]{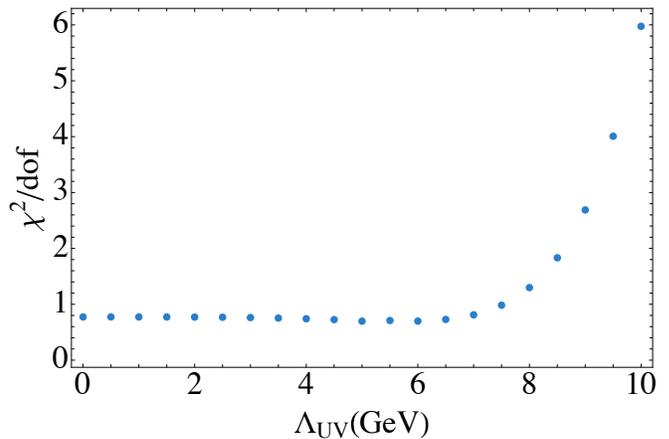}
    \caption{$\chi^2$/dof vs. $\rho_v =$ ( $\Lambda_{\rm{UV}}$ GeV)$^4$ for the best fit to the $np_{3/2}$  data \cite{Peper19}.}
    \label{fig:chisqr}
\end{figure}

\begin{figure*}[!htb]
\centering
\includegraphics[width=0.48\textwidth]{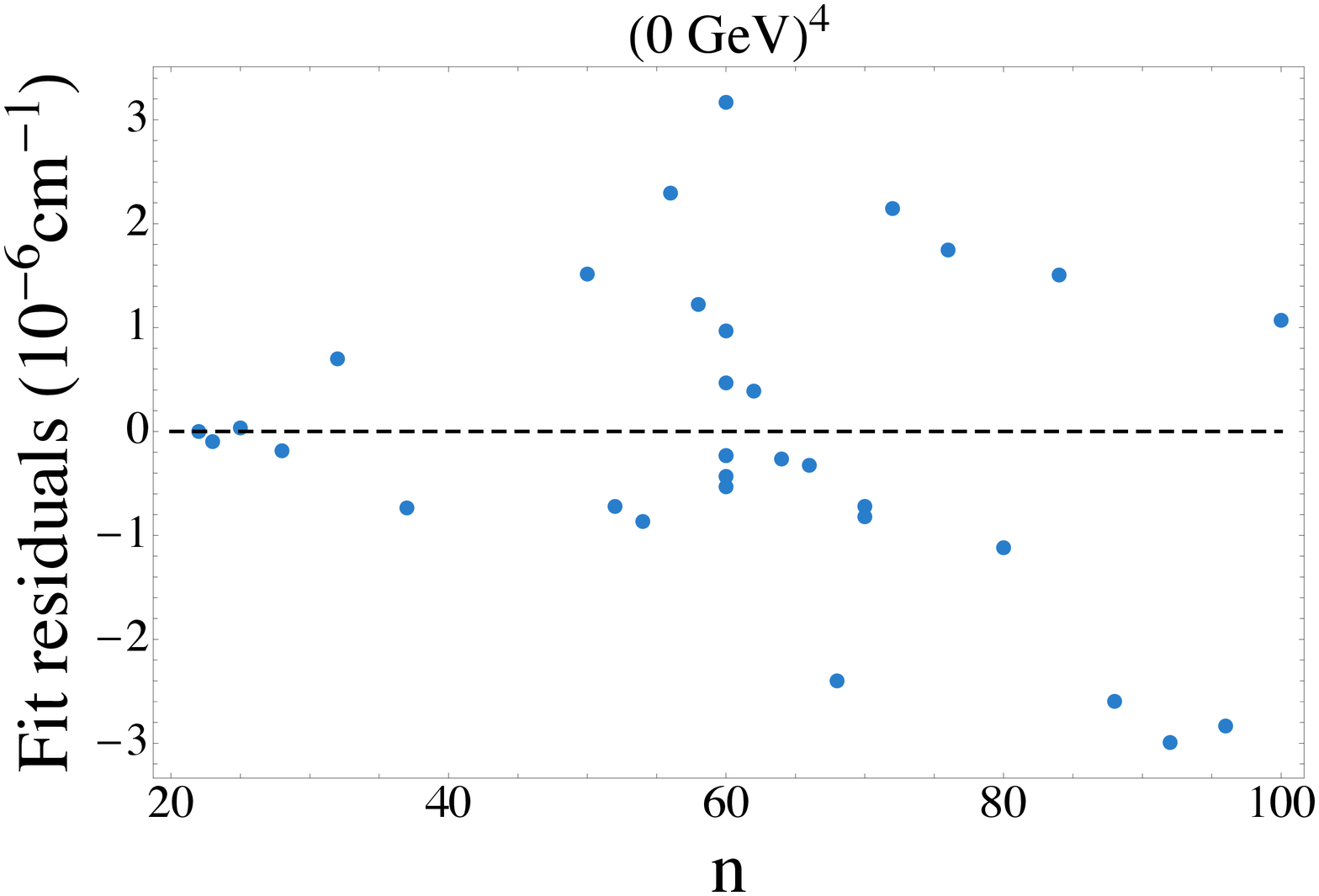}
\includegraphics[width=0.48\textwidth]{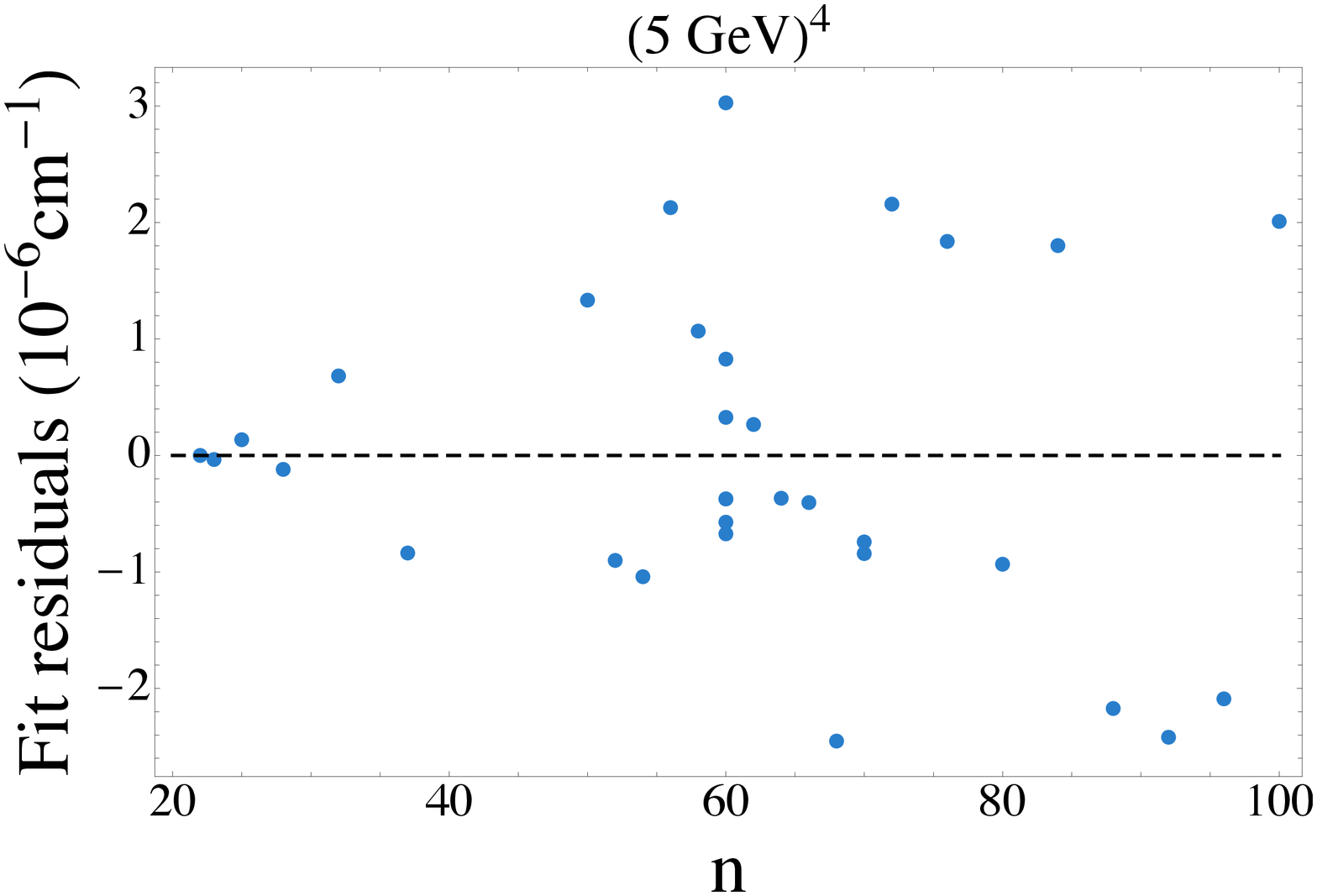}
    \newline
    \newline
    \newline
\includegraphics[width=0.48\textwidth]{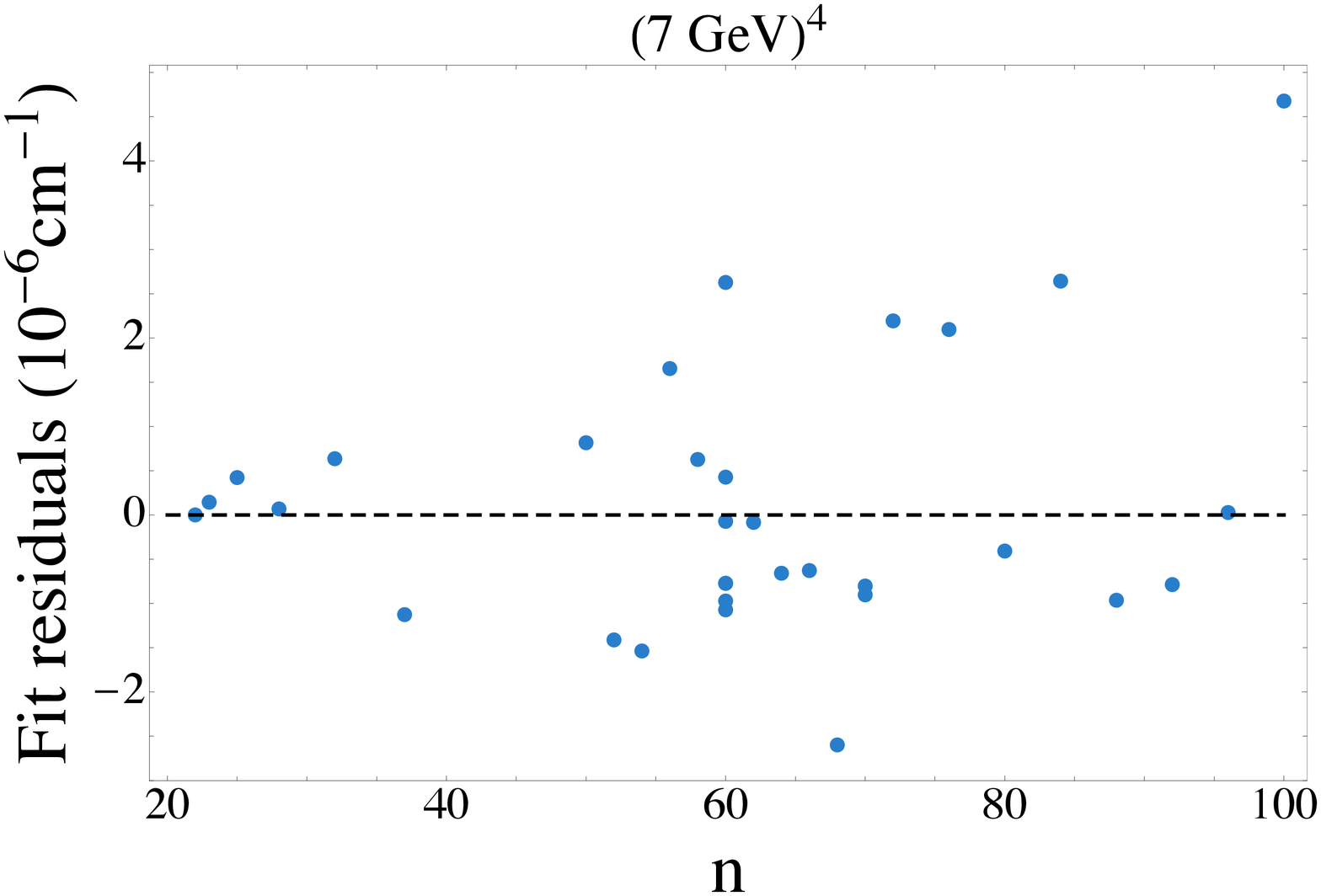}
\includegraphics[width=0.48\textwidth]{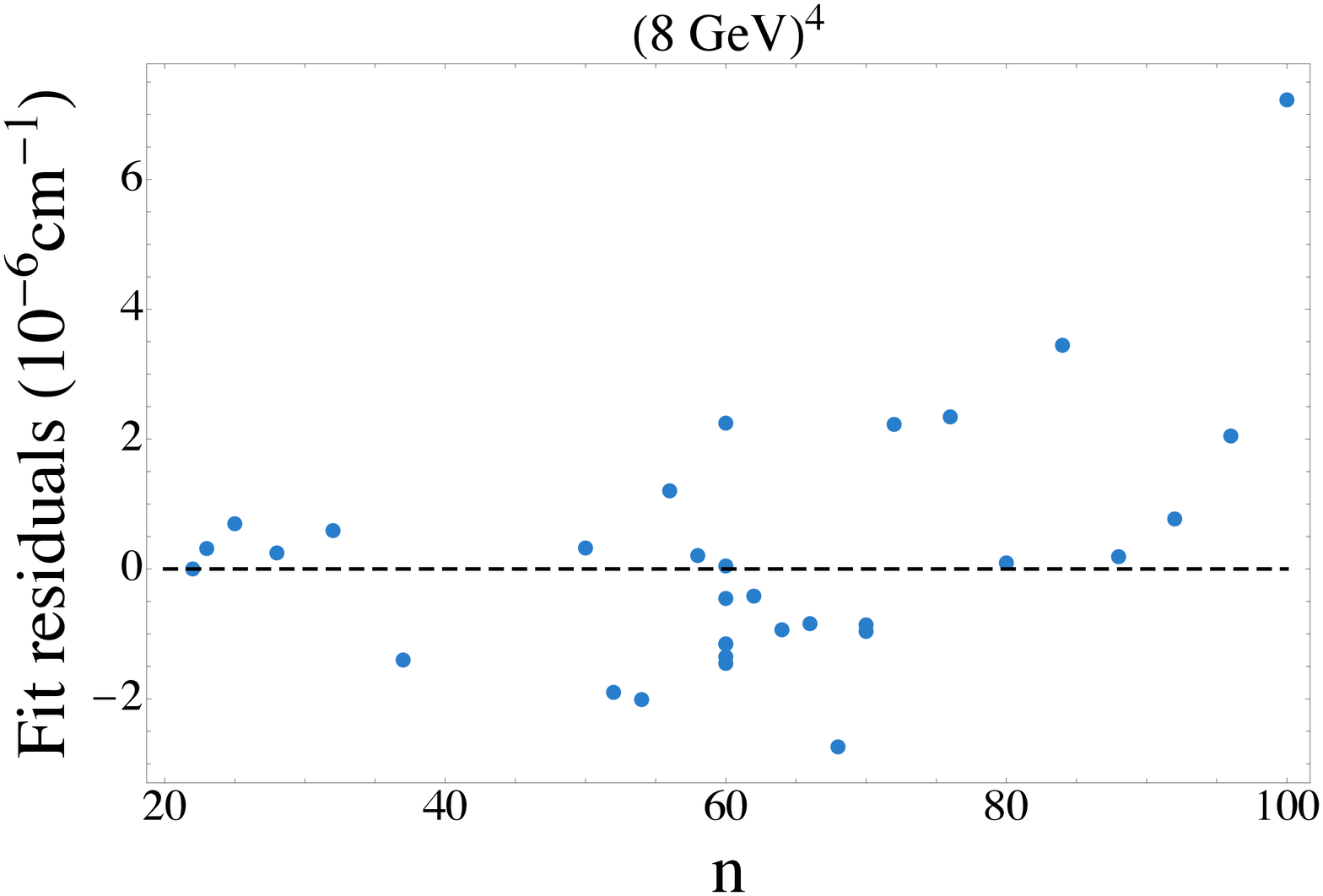}
\caption{Fit residuals for the best fit to the energy difference (w.r.t $n=22$) vs $n$ data \cite{Peper19} for selected values of the VE density $\rho_v$ (top legend of each panel).}
\label{fig:bestfit}
\end{figure*}

We first fit the energy level differences without perturbation to check that our fit was equivalent in quality to that in \cite{Peper19}, where they utilize the full Ritz formula eqn. \ref{eq:ritz}, including defects up to $\delta_6$ in eqn. \ref{defect}. After verifying this equivalence, we redid the fits including the perturbation $\delta E$. For $\rho_{v} \lesssim {(7 \ \text{GeV}})^4$ the fits with $\delta E$ are almost as good as the original Ritz formula, but rapidly deteriorate with increasing $\rho_{v}$. Table \ref{tab:table1} summarizes the fit statistics. We judge the quality of the fits by comparing the $\chi^2/$dof (degrees of freedom) in Fig. \ref{fig:chisqr} and the size and randomness of the residues of the fits with $\delta E$ to the unperturbed fits in Figure \ref{fig:bestfit}. 

Using $n_{\ast} = n-\delta_{lj}(n)$ instead of $n$ in eqn. \ref{eq:ptbformula}, makes no material difference in our results and gives an energy scale that is little changed. 

The data used in our fit of RAs extend from the $n= 22$ level to $n = 100$ corresponding to atomic sizes ranging from $10^{-8}$ to $10^{-6}$ meters. Since eqn. \ref{eq:ritzwp} is a global fit to the data it provides information on this entire distance range. Due to the $n^4$ dependence of the perturbation in eqn. \ref{eq:ptbformula}, our fit is more sensitive to large $n$, or equivalently larger atomic radii. For instance, take $\Lambda_{\rm{UV}}$ to be 7 GeV at one micron and then consider that it increases as 1/n as we probe a distance scale of 0.1 micron. We then find that, at a distance of the order of $0.1$ micron, $\Lambda_{\rm{UV}}$ is well below $100$ GeV. This rules out the predictions of the Standard Model for gravitating VE at distances of a few tenths of a micron or less.

This is our chief result, an upper limit to the scale of a gravitating vacuum $\rho_{vac} \lesssim {(7 \ \text{GeV}})^4$ at distances less than a micron  (The $7 \ \text{GeV}$ limit is valid independent of the sign of $\rho_{vac}$). Since this scale is well below the established scales of the SM, it raises the question of whether the VE gravitates at all.  There is the claim that zero-point energy does not contribute in the usual manner \cite{Donoghue21,Moreno22}, but VE is a broader concept than zero-point energy. The Higgs potential contributes to the VE and should gravitate. The absolute value of the Higgs potential is unknown, but we expect that its value should be of the same order as the difference between the local maximum of the potential and its symmetry-breaking vacuum (the difference between the top of the ``Mexican hat" and its brim), which is again of order $100$ GeV.  There are also QCD contributions to VE from chiral and gluon condensates.  Inflationary models are dependent on the gravitation of the inflaton potential. The absence of gravitating VE poses serious questions to our understanding of quantum field theory and semiclassical gravity.

\section{Validity of the Effective Field Theory description }\label{eftsec}

A recent flurry of activity, known as the UV/IR connection \cite{Castellano21,deRham22,Draper22}, explores how a consistent quantum theory of gravity could manifest itself at scales well below the Planck scale \cite{Cohen:1998zx,Hamed04}. In string theory, this is an attempt to avoid getting stuck in the swampland, but there are also attempts to understand how gravity may restrict effective field theories (EFT).  It is instructive to consider our results from the point of view of the EFT approach. 

Effective field theory is a framework that is dramatically successful when applied to the standard model (SM) of particle physics. The SM is incomplete since it does not include quantum gravity, but semiclassical gravity is easily incorporated. Following \citet{Cohen:1998zx} (CKN hereafter), we consider an EFT with highest energy scale $\Lambda_{\text{UV}}$, the cut-off beyond which the theory no longer applies. We use the EFT to describe a space-time patch, of radius $R$, with a high energy density. The maximum energy density in this EFT is $\Lambda_{\text{UV}}^4$. If this patch is large enough, it will collapse into a black hole, a state that does not appear in the EFT. Thus, the EFT must break down or change. The critical size $R_c$ for this to occur is when $R_c$ is equal to the Schwarzschild radius $R_S$ 

\begin{equation}
\begin{split}
   R_c=R_S = 2G \Lambda_{\text{UV}}^4 \frac{4 \pi}{3}  R_c^3 \sim  \Lambda_{\text{UV}}^4 R_c^3 \frac{8}{M_P^2} 
   \end{split}
 \label{eq:rlimit}
 \end{equation}
 The criterion for this NOT to happen is that its radius $R$, be less than $R_c$ i.e. 
 \begin{equation}
      R <  M_P/\sqrt{8}\Lambda_{\text{UV}}^2
      \label{eq:rbound}
 \end{equation}
imposing an IR limit to our theory, implying limitations on physical scales very different from energy cutoffs. This strongly hints that predictions of gravitating VE will depend on the distance scales considered. Alternatively, if we restrict our attention to a very large region $(R=1/\Lambda_{\text{IR}} >>1)$ then fixing $R$ restricts the EFT cutoff $\Lambda_{\text{UV}}$ to be small. If we investigate the universe and choose $R$ as today's Hubble radius, we find $\rho_{v} \sim (0.002  \ \text{eV})^4$, intriguingly close to the observed CC. This interconnection is very different from the expectations of our traditional study of quantum field theory and hints at a possible clue to solving the CC problem.

To see how the CKN bound points to VE gravitating only at short distances consider the extension of the SM to the TeV range. The TeV scale is the most popular energy scale to go beyond the SM. Supersymmetry is the most familiar of these and typically has a cutoff scale of a few to tens of TeV. We are able to calculate implications of these models, via an EFT, for gravitating VE, as long as we respect the restraint eqn. \ref{eq:rbound}.  There are no predictions for distances greater than $1-100$ microns (depending on the exact size of the cutoff in eqn. \ref{eq:rbound}) and we are not surprised that no gravitating effects of VE are observed at distances larger than microscopic.  However, below 1-10 microns the EFT predicts large effects. RA conclusively rules out this not only for TeV scales but also for all scales larger than $7$ GeV.

As an illustration of how the gravitational effects of VE with a $10$ GeV cut-off would not extend beyond microscopic distances, consider the following modified CKN bound (this modification is rather ad hoc, and we use it for illustrative purposes only)
\begin{equation}\label{modckn}
      R <  (M_P/\Lambda_{\text{UV}})^{\alpha} \Lambda_{\text{UV}}^{-1}
\end{equation}
Similar forms of the CKN bound were used in \cite{Hsu04,Rama21}. If we set $\alpha= 1/2$ in eqn. \ref{modckn} as the true UV/IR restriction, the SM EFT does not make predictions for $\Lambda_{\text{UV}} \sim 10 \ \text{GeV}$ at distances greater than a micron. Thus, we would not expect to observe VE gravitating at macroscopic distances. However, at distances less than $1$ micron the EFT for the SM would be predictive, and RA data, as we have seen, rule out VE gravitating at $10$ GeV energy scales. Eqn. \ref{modckn} is an example of an ansatz  distinguishing between the gravitating effects of VE at large and small distances. RAs do not see any evidence that this or any VE gravitates.

\vspace{0.3cm}

\section{Conclusion}
RAs have a long history of utility in studies of atomic physics and chemistry, and recently in quantum computing.  We claim that fundamental physics and cosmology can be added to the list. The idea that atomic physics experiments are capable of providing any information about the expansion of the universe is amazing and a celebration of the unity of physics. Although Rydberg measurements of the CC will never replace conventional astronomical measurements, they provide further evidence for the need to better understand the origin of the CC. What is remarkable is not that RAs are a good way to measure the CC but that they can do it at all!

The place where RAs add unprecedented information is in their ability to measure local effects of VE. We find that precision studies of RAs put an upper limit on the gravitating VE of $\rho_{vac} \lesssim {(7 \ \text{GeV}})^4$ at distance scales less than a micron. This is well below the typical energy scale of the SM and so contradicts our naive expectations.  It points to a shutting off of gravitating VE at all scales as the path to understand the smallness of the CC on cosmological scales.

The limit to the precision of energy levels attainable in RA is the line width \citep{Song22}, so it may be possible to consider an improvement by a factor of $\sim 10$ for $E_n$ when $n = 100$. This improves as $(n/100) ^{1.6}$ as we increase $n$.  Such improvements could lead to both an improved upper limit on the scale of gravitating VE and tighter limits at distances significantly shorter than a micron. As the precision of RAs continues to increase and as higher excited states are studied, we may someday probe QCD scales ($ \sim 1$ GeV)  and/or improve restrictions of gravitating VE at distances less than 0.1 micron, leading to a better understanding of how gravity can be incorporated into our quantum field-theoretic descriptions of nature. 

\section{Acknowledgement}
We thank S. Watson, J. Hubisz, S. Ballmer, and members of the 316 seminar series for helpful discussions. We are grateful to F. Merkt and  G. Clausen for illuminating discussions. We also thank S. L. Adler for useful comments on our preprint \cite{Adler22}. We thank an anonymous referee for stressing to us that macro and mesoscopic experience already puts stringent restrictions on the CC. We also thank an anonymous referee for bringing reference \cite{Gill87} to our attention. SKK acknowledges support from the National Science Foundation through grant AST-2006684. AP acknowledges support from the US Department of Energy (DOE), Office of Science, Office of High Energy Physics under Award Number {DE-SC0009998}.

\begin{table*} [!htbp]
    \centering
  \begin{ruledtabular}
  \begin{tabular}{lcccc}
   $\rho_v$ & $(0 \  \text{GeV})^4$  &$(5 \ \text{GeV})^4 $ & $(7 \ \text{GeV})^4 $ &$(8 \ \text{GeV})^4$
  \\
  \colrule
  $\delta_0$ & 1.7108778(1) & 1.7108779(1) & 1.7108781(1) & 1.7108782(1)\\ 
  $\delta_2$ & 0.2330963(4) & 0.2330741(4) & 0.2330110(7) & 0.2329508(5) \\
  $\chi^2/dof$ & 0.7731793 & 0.7093683 & 0.8110845 & 1.2986782
\end{tabular}
\end{ruledtabular}
    \caption{Fit parameters for the best fit to the energy difference (w.r.t $n=22$) vs $n$ data \cite{Peper19} for selected values of the VE density $\rho_v$.}
    \label{tab:table1}
\end{table*}

\bibliography{ryd.bib}

\end{document}